\documentclass[aps,prl,preprint,groupaddress]{revtex4}
\usepackage{graphicx}
\usepackage{dcolumn}
\usepackage{bm}
\usepackage{color}
\usepackage{amsmath}
\begin{document}
%
\title{Large Enhancement of Thermoelectric Efficiency Due to a Pressure-Induced Lifshitz Transition in SnSe}
%
\author{T. Nishimura$^{1}$}
\author{H. Sakai$^{1,2}$}\email[Corresponding author: ]{sakai@phys.sci.osaka-u.ac.jp}
\author{H. Mori$^{1}$}
\author{K. Akiba$^{3}$}
\author{H. Usui$^{1}$}
\author{M. Ochi$^{1}$}
\author{K. Kuroki$^{1}$}
\author{A. Miyake$^{3}$}
\author{M. Tokunaga$^{3}$}
\author{Y. Uwatoko$^{3}$}
\author{K. Katayama$^{1}$}
\author{H. Murakawa$^{1}$}
\author{N. Hanasaki$^{1}$}
\affiliation{$^1$Department of Physics, Osaka University, Toyonaka, Osaka 560-0043, Japan.\\
$^2$PRESTO, Japan Science and Technology Agency, Kawaguchi, Saitama 332-0012, Japan.\\
$^3$The Institute for Solid State Physics, University of Tokyo, Kashiwa 277-8581, Japan}
%
\begin{abstract}
Lifshitz transition, a change in Fermi surface topology, is likely to greatly influence exotic correlated phenomena in solids, such as high-temperature superconductivity and complex magnetism.
However, since the observation of Fermi surfaces is generally difficult in the strongly correlated systems, a direct link between the Lifshitz transition and quantum phenomena has been elusive so far.
Here, we report a marked impact of the pressure-induced Lifshitz transition on thermoelectric performance for SnSe, a promising thermoelectric material without strong electron correlation.
By applying pressure up to 1.6 GPa, we have observed a large enhancement of thermoelectric power factor by more than 100\% over a wide temperature range (10-300 K).
Furthermore, the high carrier mobility enables the detection of quantum oscillations of resistivity, revealing the emergence of new Fermi pockets at $\sim$0.86 GPa.
The observed thermoelectric properties linked to the multi-valley band structure are quantitatively reproduced by first-principles calculations, providing novel insight into designing the SnSe-related materials for potential valleytronic as well as thermoelectric applications.
\end{abstract}
%
%
\maketitle
%
In 1960's, Lifshitz proposed an electronic phase transition that has no relevant Landau-type order parameter accompanied by symmetry breaking\cite{Lifshitz1960}.
This so-called Lifshitz transition is described by the collapse or emergence of Fermi surfaces caused by external parameters, identified as a topological transition in condensed matter physics\cite{Lifshitz1960,Volvic2017LowTPhys}.
Effects of the Lifshitz transition on physical properties have been long investigated; at early stages they were discernible but not large, as exemplified by the superconductivity in simple metals\cite{Chu1970PRB,Watlington1977PRB}.
More recently, it was suggested that the Lifshitz transition plays a significant role in strongly correlated systems, such as high-$T_c$ superconductors\cite{Norman2010PRB,Liu2010NatPhys} and heavy-fermion itinerant magnets\cite{Sandeman2003PRL,Yamaji2007JPSJ,Yelland2011NatPhys}.
However, the strong coupling with various degrees of freedom (e.g., charge, spin and lattice) tends to complicate the situation and hence masks the genuine role of the Lifshitz transition in quantum phenomena.
Furthermore, the direct observation of Fermi surface in strongly correlated system has been an experimental challenge, which makes it virtually impossible to precisely detect the Lifshitz transition.
%
\par
\begin{figure}
\begin{center}
\includegraphics[width=.75\linewidth]{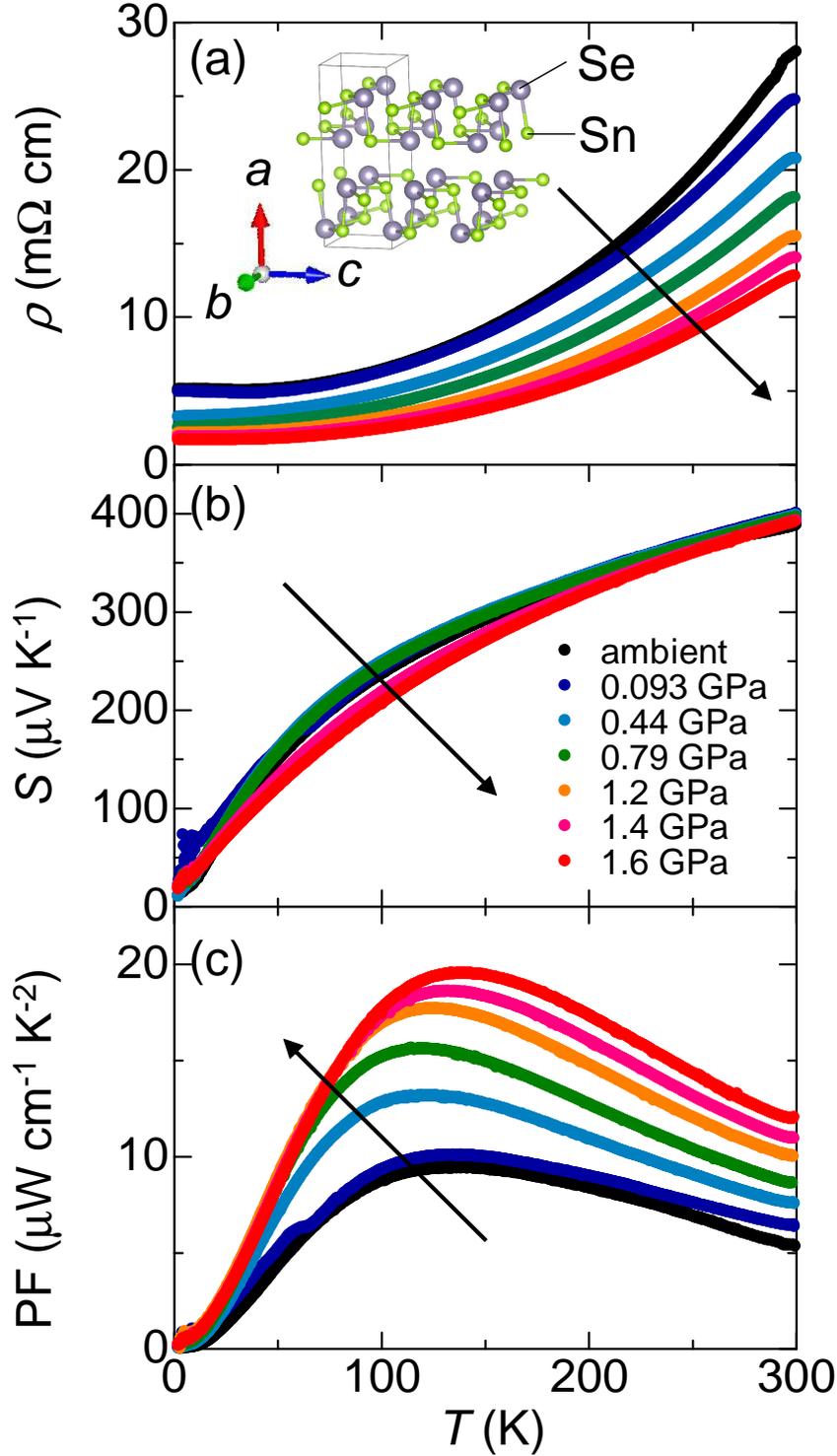}
\end{center}
\caption{\label{fig:RT_ST}(color online) Temperature profiles of in-plane (a) resistivity $\rho$, (b) thermopower $S$, and (c) power factor ${\rm PF}\!=\!S^2/\rho$ for a SnSe single crystal at various pressures up to 1.6 GPa. The electrical and heat currents were applied within the $bc$ plane. Inset: Crystal structure of orthorhombic SnSe ($Pnma$), consisting of black-phosphorus-type buckled layers.
}
\end{figure}

%
Lifshitz transition is accompanied by anomalies in density of states (i.e., van Hove singularity), which is anticipated to affect not only quantum ordered phases but also transport phenomena\cite{Barber2018PRL}.
Among them, thermopower is one of the most sensitive transport coefficients for the Fermi surface variation\cite{Okamoto2010PRB}, since it is proportional to the energy derivative of density of states within a semiclassical Boltzmann theory.
Thus, thermoelectric materials can be promising for clarifying the pure impact of the Lifshitz transition.
As a candidate for this, we focus on SnSe, which has very recently attracted significant attention due to the high thermoelectric figure of merit at around 900 K~\cite{Zhao2014Nature}.
In addition to the ultra-low thermal conductivity, the multi-valley valence bands have been suggested as an origin of the excellent performance\cite{Kutorasinski2015PRB, Guo2015PRB, Mori2017PRB, Zhao2016Science, Wang2018NatCom, Nagayama2018JJAP, Pletikosic2018PRL}.
Therefore, we here aimed to drive the Lifshitz transition by employing hydrostatic pressure to continuously tune the relative band-filling of the multi-valley states for SnSe.
Another advantage of SnSe is its high carrier mobility due to the weak electron correlation, which reaches $\sim$2000 cm$^2$V$^{-1}$s$^{-1}$ at low temperatures\cite{Maier1997JEelectronMater}, comparable to that for black phosphorus\cite{Li2014NatNano} with the similar buckled honeycomb layered structure [inset to Fig. \ref{fig:RT_ST}(a)].
This enabled us to directly reveal the Fermi surface change via the quantum oscillation observable over the applied pressure range.
%
\par
%
The hole ($p$)-type carrier concentration $n_h$ in SnSe varies by more than two order of magnitude by the control of chemical defects\cite{Wang2018NatCom,Yamashita2018JPSJ} and/or intentional chemical substitution\cite{Zhao2016Science}.
For instance, pristine degenerate-semiconductor-like SnSe exhibits $n_h\!\sim\! 3\!\times\! 10^{17}$ cm$^{-3}$ at room temperature\cite{Zhao2014Nature}, whereas $n_h$ reaches $\sim 4\!\times\! 10^{19}$ cm$^{-3}$ for the Na-doped metallic SnSe\cite{Zhao2016Science}.
In the present study, we have adopted moderately hole-doped SnSe with $n_h\!\sim\! 2\!\times\! 10^{18}$ cm$^{-3}$ (for estimation, see below), which shows nice metallic behavior down to the lowest temperature [Fig. \ref{fig:RT_ST}(a)].
For this carrier concentration, the Fermi surfaces consist of two tiny hole pockets along the $\Gamma$-Z line at ambient pressure [Fig. \ref{fig:theory}(a)], which provides an ideal platform for detecting changes in the Fermi surface topology. 
In the following, we will show pressure-induced Lifshitz transition in SnSe, across which the thermoelectric efficiency is largely enhanced by the emergence of new hole pockets.
%
\par
%
SnSe single crystals were synthesized by a Bridgman method.
Transport measurements were performed by using a piston-cylinder-type high-pressure cell in a Physical Properties Measurement System (Quantum Design)~\cite{SM}.
First-principles calculations were performed using the VASP code~\cite{VASP1,VASP2} with the  PBEsol-GGA functional~\cite{PBE,PBEsol} and the projector augmented wave method~\cite{PAWmethod} including spin-orbit coupling on a $6\times14\times14$ $\bm{k}$-mesh sampling of the Brillouin zone.
The plane wave energy cutoff was 400 eV.
We determined the crystal structures through structural optimization under the hydrostatic pressures of 0, 0.5, 1, 1.5, and 2 GPa assuming them to be orthorhombic (space group: $Pnma$), and then we performed the band calculations using these structures~\cite{SM}.
%
\par
%
%
We first show the pressure variation of thermoelectric properties for single-crystalline SnSe.
With increasing pressure, the in-plane resistivity ($\rho$) progressively decreases while keeping the metallic behavior [Fig. 1(a)], as is usually observed for metals and degenerate semiconductors\cite{note_semi}.
The $\rho$ value at 1.6 GPa is less than half of that at ambient pressure irrespective of temperature.
On the other hand, the temperature profile of thermopower ($S$) shows anomalous pressure dependence; the $S$ value is nearly constant against pressure over the entire measured temperature regime [Fig. 1(b)].
The reduction in $S$ at 1.6 GPa is barely discernible at 50-150 K, which is, at most, only $\sim$10\% of $S$ at ambient pressure.
Consequently, the thermoelectric power factor, PF$=S^2/\rho$, exhibits a large enhancement upon pressure.
The PF value peaks around 150 K, which reaches $\sim$20 $\mu$W/cm K$^2$, more than double of that at ambient pressure and close to the maximum value of the Na-doped SnSe with one order of magnitude higher carrier concentration\cite{Zhao2016Science}.
Considering the observed large decrease in $\rho$ with pressure, the origin of the PF enhancement may be sought in the pressure-induced increase in bandwidth, i.e., decrease in effective mass.
However, this should lead to a significant decrease in $S$, as $S$ is proportional to the effective mass assuming a free-electron model, which is inconsistent with the experimental results.
%
\par
\begin{figure}
\begin{center}
\includegraphics[width=\linewidth]{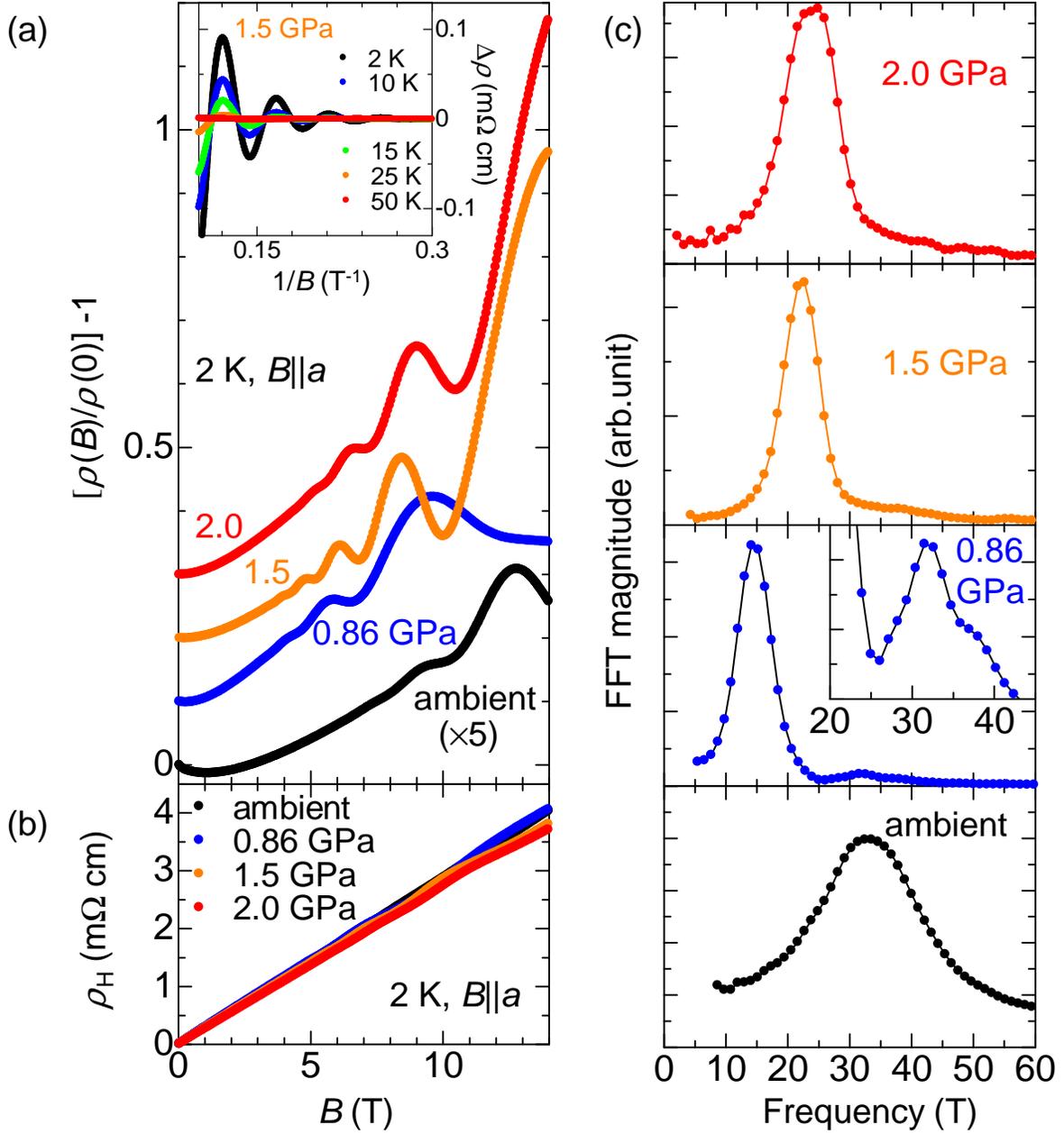}
\end{center}
\caption{\label{fig:SdH}(color online) Field ($B$) dependence of (a) magneto-resistivity $\rho(B)/\rho(0)-1$ and (b) Hall resistivity $\rho_{\rm H}$ along the $bc$ plane at various pressures at 2 K. The magnetic field was applied along the $a$ axis (out of plane). For clarity, each curve of the magneto-resistivity is shifted vertically by 0.1 in panel (a). Inset: temperature dependence of oscillatory part of $\rho$ at 1.5 GPa plotted versus $1/B$. As typical behavior of the SdH oscillation, the amplitude is progressively suppressed with increasing temperature.
(c) Plots of fast Fourier transform of the oscillatory part of $\rho$ at 2 K at various pressures.
} 
\end{figure}
\begin{figure}
\begin{center}
\includegraphics[width=\linewidth]{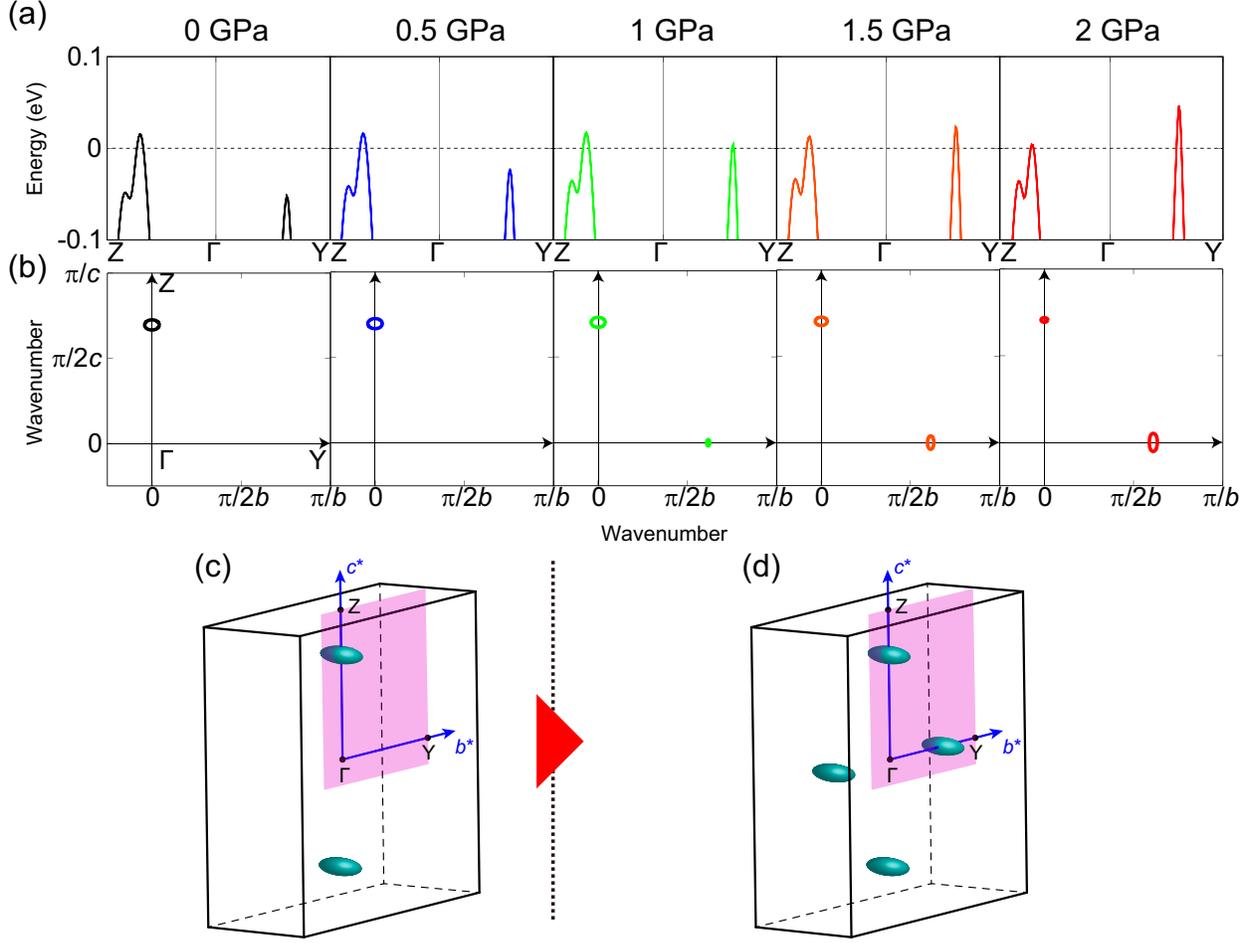}
\end{center}
\caption{\label{fig:theory}(color online) Pressure variation of (a) electronic band structures and (b) Fermi surfaces on the $k_a=0$ plane. The Fermi level, which is set to zero in panel (a), was determined to provide the fixed carrier density $n_h=4\times 10^{-4}$ $e/\rm{unit\ cell}$ for each pressure. This value corresponds to $n_h\!=\! 1.9\!\times\! 10^{18}$ cm$^{-3}$ for the unit cell volume at 0 GPa, nearly agreeing with the carrier density estimated from the experimental Hall coefficient ($2.3\!\times\! 10^{18}$ cm$^{-3}$). 
Schematic illustration of the Fermi surfaces exhibiting the Lifshitz transition from (c) a two-valley state to (d) a four-valley one. The shaded plane corresponds to the plotted area in panel (b).
}
\end{figure}

%
To reveal the pressure-induced variation in electronic structure for SnSe, we have measured the magneto-resistance and Hall effects at 2 K at various pressures.
As shown in Fig. \ref{fig:SdH}(b), the field-linear Hall resistivity remains almost unchanged up to 2.0 GPa, implying nearly constant carrier density with pressure.
On the other hand, as shown in Fig. \ref{fig:SdH}(a), the magneto-resistivity markedly depends on pressure, where clear Shubnikov-de Haas (SdH) oscillations are superimposed to the non-oscillatory positive magneto-resistance effect.
The magnitude of the background positive magneto-resistance effect is strongly enhanced as the pressure increases; the value of $\rho(B)/\rho(0)-1$ is less than 0.01 at $B\! =\!5$ T at ambient pressure, while it exceeds 0.1 at 2.0 GPa.
Furthermore, the period and amplitude of the SdH oscillation clearly varies upon pressure.
To quantitatively evaluate the pressure dependence, we have extracted the oscillatory component of magneto-resistivity $\Delta \rho$ [see inset to Fig. \ref{fig:SdH}(a)] and performed the fast Fourier transform with respect to $1/B$.
At ambient pressure, a main broad peak is visible at 33 T in the Fourier transform [Fig. \ref{fig:SdH}(c)], leading to the Fermi surface area of 0.31 nm$^{-2}$ in the reciprocal space.
Assuming that two equivalent weakly-anisotropic 3D Fermi pockets are located along the $\Gamma$-Z line in the Brillouin zone\cite{Zhao2016Science}[see Fig. \ref{fig:theory}(c)], this extremal cross section leads to $n_h\! =\! 2.7\!\times\! 10^{18}$ cm$^{-3}$~\cite{note}, which nicely coincides with the value estimated from the Hall coefficient ($n_h\! =\! 2.3\!\times\! 10^{18}$ cm$^{-3}$).
By applying the pressure of as small as 0.86 GPa, intriguingly, the Fourier transform dramatically changes; a sharp main peak manifests itself at 15 T and the peak observed at ambient pressure remains at $\sim$32 T as a small secondary peak.
With further increasing pressure to 1.5 GPa, the main peak rapidly moves to higher frequency and almost merges with the secondary one at $\sim$22 T.
Above 1.5 GPa, the frequency of the main peak gradually increases, reaching $\sim$24 T at 2.0 GPa.
The observed pressure variation in SdH oscillations clearly indicates a pressure-induced Lifshitz transition; a new Fermi pocket emerges at around 0.86 GPa (corresponding to the peak at 15 T), which further evolves in size with increasing pressure.
The increase in the number of valleys in the Brillouin zone across the Lifshitz transition is advantageous to achieving both high thermopower and high electrical conductivity\cite{Mori2017PRB}. 
To show the detailed relation between the variation in band structure and the enhanced thermoelectric efficiency, we shall compare the experimental data to the results of first-principles calculation.
%
%
\par
\begin{figure}
\begin{center}
\includegraphics[width=.55\linewidth]{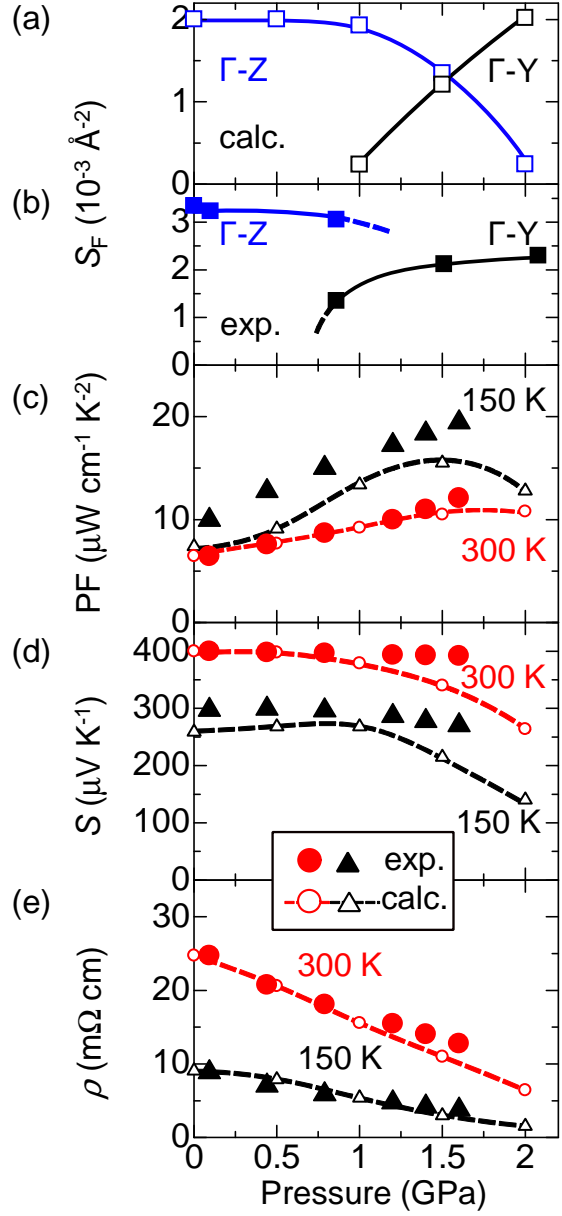}
\end{center}
\caption{\label{fig:comparison}(color online) (a) Calculated and (b) experimental values of extremal cross section of Fermi surfaces perpendicular to the $a$ axis, $S_{\rm F}$, versus pressure. The former is the result from the first-principles calculations, while the latter is estimated from the SdH frequencies. The corresponding (c) PF, (d) $S$, and (e) $\rho$ along the $bc$ plane at 300 K and 150 K are also plotted. The experimental (theoretical) results are denoted by closed (open) symbols.
For calculating $\rho$, we assume a constant relaxation time independent of pressure at each temperature, which is determined to be 1.5$\times$10$^{-14}$ s (4.6$\times$10$^{-14}$ s) at 300 K (150 K) so as to reproduce the experimental $\rho$ value at ambient pressure. We also assume a common relaxation time irrespective of the valence bands.
The direction of the current and thermal gradient is set along the $b$ axis in the calculations.
}
\end{figure}

%
Figure \ref{fig:theory}(a) shows the calculated band structures in the vicinity of the Fermi energy along the high-symmetry $\Gamma$-Y and $\Gamma$-Z lines at selected pressures [For the band structure in a larger energy scale, see the supplementary information (Fig. S1) and Ref. \onlinecite{Zhang2016JMaterChem}].
The corresponding Fermi surfaces on the momentum plane parallel to the layers of SnSe ($k_a\! =\! 0$) are shown in Fig. \ref{fig:theory}(b).
At ambient pressure (0 GPa), the valence band along the $\Gamma$-Z line crosses the Fermi energy, resulting in two small hole pockets in the entire Brillouin zone, as schematically shown in Fig. \ref{fig:theory}(c).
With increasing pressure, the top of the valence band along the $\Gamma$-Y line progressively rises and crosses the Fermi energy at around 1 GPa.
Consequently, new hole pockets emerge along the $\Gamma$-Y line, leading to a four-valley valence band structure [Fig. \ref{fig:theory}(d)].
Figures \ref{fig:comparison}(a) and (b) compare the theoretical and experimental pressure variations of these hole pockets, where the cross section ($S_F$) on the $k_a\!=\!0$ plane deduced from the first-principles calculation and experimental SdH frequency are plotted versus pressure respectively.
The calculated $S_F$ value of the hole pocket along the $\Gamma$-Z line is nearly constant up to 1 GPa, above which $S_F$ steeply decreases with increasing pressure as the hole pocket along the $\Gamma$-Y emerges and evolves.
As a result, two pockets are comparable in size at 1.5 GPa [Fig. \ref{fig:comparison}(a)].
This behavior semi-quantitatively reproduces the pressure dependence of experimental $S_F$ shown in Fig. \ref{fig:comparison}(b).
At around 2 GPa, the $S_F$ value for the pocket along the $\Gamma$-Y is much larger than that for the $\Gamma$-Z in calculation, which is unlikely to be consistent with experiment; we detected a single SdH frequency slightly larger than that at 1.5 GPa without clear splitting of $S_F$ [Fig. \ref{fig:comparison}(b)], indicating that two hole pockets still have similar $S_F$ values.
Except for such overestimation at the high-pressure region, the prssure-induced change of Fermi surface topology is essentially reproduced by the first-principles calculation.
%
%
\par
%
For the comparison of thermoelectric features, we present the calculated PF, $S$, and $\rho$ values together with the experimental values (at 300 K and 150 K) in Figs. \ref{fig:comparison}(c), (d) and (e), respectively.
The overall pressure dependence of the theoretical and experimental results shows quantitative agreement regardless of temperature.
In particular, the monotonic increase in PF with pressure up to 1.5 GPa is well reproduced by the calculation, which results from the significant decrease in $\rho$ [Fig. \ref{fig:comparison}(e)] while keeping nearly constant $S$ [Fig. \ref{fig:comparison}(d)].
The calculated PF value reaches the maximum at around 1.5 GPa, where the system exhibits the four-valley band structure.
Thus, the pressure-induced increase in the number of valence band valleys is relevant to the enhanced thermoelectric efficiency\cite{note_decomposition}.
With further increasing pressure, the calculated PF begins to decrease owing to the large reduction in $S$, which corresponds to the approach to the two-valley state with the hole pockets along the $\Gamma$-Y line.
In the present experiment, however, the maximum pressure in the thermopower measurements was 1.6 GPa, up to which the PF value continues to increase with pressure.
We should note here that the calculation tends to underestimate the $S$ value above 1.4 GPa, which may indicate that the four-valley state, i.e., the hole pocket along the $\Gamma$-Z line, is in reality robust even above 1.5 GPa, as is suggested from the SdH experiments.
%
%
\par
%
Interestingly, a recent first-principles calculation revealed that the band structure for SnSe predicted from the crystal structure at 790 K exhibits four-valley character as well, owing to the temperature-induced rise of the valence band top along the $\Gamma$-Y~\cite{Mori2017PRB}.
Although increasing temperature and applying pressure have an opposite impact on the lattice size, it should be noted that both temperature and pressure similarly increase the $b/c$ value towards unity (see supplementary Fig. S7 and Refs. \onlinecite{Chattopadhyay1986JPhysChemSol,Loa2015JPhysC,Zhang2015Nanoscale}) and hence tend to stabilise the higher-symmetry $Cmcm$ phase\cite{Alptekin2011JMM,Yu2016SR,Cong2018JPCC,Loa2015JPhysC,Zhang2015Nanoscale}.
(Note here that the structural transition from $Pnma$ to $Cmcm$ occurs at $\sim$10 GPa\cite{Loa2015JPhysC,Zhang2015Nanoscale,Chen2017PRB}, much higher than the present pressure range.)
Thus, the $b/c$ value can be a key factor for raising the energy of the valence band along the $\Gamma$-Y line and achieving the four-valley structure.
%
\par
%
In conclusion, we have experimentally and theoretically revealed a pressure-induced Lifshitz transition of the Fermi surfaces for a high-performance thermoelectric SnSe.
By tuning the external pressure, a four-valley band structure is realized, which leads to the enhancement of thermoelectric power factor by more than 100\% over a wide temperature range (10-300 K).
The pressure-sensitive Fermi surface topology has been here directly evidenced by observing the SdH oscillation.
It may be of interest as a future research to study the corresponding change in band structure by means of optical spectroscopy\cite{Xi2013PRL,Martin2014PRB}.
The present result demonstrates the multi-valley state is of great importance for achieving high thermoelectric performance, providing a guide for band engineering for the SnSe-based thermoelectrics.
Furthermore, since atomically thin SnSe films have recently attracted much attention as a new 2D material\cite{Pletikosic2018PRL,Li2013JACS}, the controllability of valley structure presented here will help for exploring their novel functions in the emerging valleytronics.
%
\begin{acknowledgements}
The authors are grateful to H. Takahashi and S. Tsuchihashi for experimental advices and fruitful discussions.
The work was in part supported by the JST PRESTO (Grant No. JPMJPR16R2), the JST CREST (Grant No. JPMJCR16Q6), the JSPS KAKENHI (Grant Nos. JP16H06015, JP17K14108, JP16K13838), and the Asahi Glass Foundation.
\end{acknowledgements}
%

%

\begin{thebibliography}{}
\bibitem{note_decomposition} The contribution of the new hole pocket ($\Gamma$-Y) to $\rho$ and $S$ is theoretically clarified by calculating the valley-decomposed transport coefficient tensors (see Fig. S4 for details).
\end{thebibliography}
\end{document}